\documentclass[%
 aip, pop, citeautoscript,
 amsmath,amssymb,
 reprint,%
]{revtex4-2}

\usepackage{graphicx}
\usepackage{dcolumn}
\usepackage{bm}

\usepackage[utf8]{inputenc}
\usepackage[T1]{fontenc}
\usepackage{mathptmx}
\usepackage{etoolbox}
\usepackage{multirow}

\usepackage{float}

\usepackage{color}

\makeatletter
\def\@email#1#2{
 \endgroup
 \patchcmd{\titleblock@produce}
  {\frontmatter@RRAPformat}
  {\frontmatter@RRAPformat{\produce@RRAP{*#1\href{mailto:#2}{#2}}}\frontmatter@RRAPformat}
  {}{}
}

\usepackage[]{hyperref}
\usepackage{xcolor}
\definecolor{AIPBlue}{RGB}{61, 180, 229}
\hypersetup{
    colorlinks,
    linkcolor = AIPBlue,
    citecolor = AIPBlue,
    urlcolor = AIPBlue
}

\makeatother
\begin{document}

\preprint{AIP/123-QED}

\title[Physics of Fluids]{Optimal standoff distance for a highly focused microjet penetrating a soft material}

\author{Daichi Igarashi}
\author{Kento Kimura}
\author{Nanami Endo}
\author{Yuto Yokoyama}
\author{Hiroaki Kusuno}
\author{Yoshiyuki Tagawa}
 \email[Author to whom correspondence should be addressed: ]{tagawayo@cc.tuat.ac.jp}
 \altaffiliation{Also at Institute of Global Innovation Research, Tokyo University of Agriculture and Technology, Koganei, Tokyo 184-8588, Japan}

\affiliation{Department of Mechanical Systems Engineering, Tokyo University of Agriculture and Technology, Koganei, Tokyo 184-8588, Japan}

\begin{abstract}
A needle-free injector using a highly focused microjet has the potential to minimize the invasiveness of drug delivery.
In this study, the jet penetration depth in a soft material---which is a critical parameter for practical needle-free injections---was investigated.
We conducted jet penetration experiments by varying the inner diameter of the injection tube and the standoff distance between the meniscus surface and the soft material.
Interestingly, the results showed that the penetration depths peaked at certain distances from the meniscus, and the positions shifted further away as the inner diameter was increased.
By analyzing the velocity distribution of the microjet, the peak positions of the penetration depth and the maximum velocities were inconsistent due to the effects of the jet shape.
To account for this, we introduce the concept of the ``jet pressure impulse,'' a physical quantity that unifies the velocity and jet shape.
However, direct estimation of this parameter from experimental data is challenging due to limitations in spatiotemporal resolution.
Therefore, we used numerical simulations to replicate the experimental conditions and calculate the jet pressure impulse.
Remarkably, the results show that the jet pressure impulse has peak values, which is consistent with the penetration depth.
In addition, there is a correlation between the magnitude of the jet pressure impulse and the penetration depth, highlighting its importance as a key parameter.
This study underlines the importance of the jet pressure impulse in controlling the penetration depth of a focused microjet, providing valuable insights for the practical use of needle-free injection techniques.
\end{abstract}

\maketitle

\section{Introduction}\label{sec:Intro}
Needle-based drug delivery offers the convenience of precisely conveying a drug to the desired site at the required depth.
However, needle-based injections can be problematic due to the potential for the spread of infection from needle reuse\cite{jagger1988rates, kane1999transmission, shah1996detection, kermode2004unsafe} and some people's fear of needles\cite{taddio2012survey, wright2009fear, mcmurtry2015far}.
A needle-free injector\cite{giudice2006needle, mitragotri2005immunization, robles2020soft, huang2022effect} has been developed to address these issues; this can inject a drug directly into the body as a liquid jet, and it is expected that this approach could address the inherent disadvantages of needle injection.
Various kinds of needle-free injectors have been developed, including those using a spring\cite{baker1999fluid, schoubben2015dynamic} or compressed gas\cite{mohizin2020effect, portaro2015experiments} as the driving force to generate a liquid jet.
The liquid jets used in commercial needle-free injectors have a diffused shape in which the jet diameter at the tip is larger than the nozzle diameter.
A liquid jet with a diffused shape has a large contact area with the skin; the vertical stress it applies to the skin is thus large, and this can potentially cause tissue damage \cite{cu2020delivery, baxter2005jet, mitragotri2006current}.
Therefore, the development of a minimally invasive needle-free injector that causes less damage and pain is desirable.

To this end, a highly focused supersonic liquid microjet is being developed by Tagawa \textit{et~al.}\cite{tagawa2012highly} This device ejects a liquid jet with a focused shape; this is achieved by making the air--liquid interface concave, taking advantage of the flow-focusing effect.
In addition, much faster jets can be generated by using a laser as the driving force.
It is expected that this microjet will not only reduce pain but also increase the penetration efficiency and controllability of the jet\cite{tagawa2013needle, moradiafrapoli2017high, rohilla2020feasibility, krizek2020needle}.
Successful measurements of stress fields during jet penetration using this device have been made,\cite{miyazaki2021dynamic} and it has also been used to penetrate rat skin\cite{kiyama2019visualization}; as a result, the penetration phenomenon is becoming better understood.
However, it has not yet been put into practical use because the factors affecting the penetration depth have not yet been elucidated.
These must be controlled according to the type and purpose of the drug-delivery system.

Jet-injection experiments (described in Sec.~\ref{sec:experiment}) suggest that the jet velocity alone is insufficient for determining the penetration depth, as reported in previous studies\cite{baxter2006needle, taberner2012needle}.
Other studies have attempted to explain the penetration depth in terms of impact forces or pressures, but these have not provided a fully satisfactory explanation
\cite{shergold2006penetration, stachowiak2007piezoelectric, rohilla2020loading, rohilla2019vitro}. 
To resolve this issue, we expect that the shape of the jet must also be considered to find the parameters that affect the penetration depth. 
This study sought to elucidate a key parameter determining the penetration depth by focusing on the jet pressure impulse, a physical quantity that unifies the velocity and shape of the jet.
The jet pressure impulse is defined as the momentum passing through a unit cross-sectional area per unit time. 
Attempts were made in this work to obtain these values experimentally, but the jet pressure impulse was difficult to estimate due to the limited spatiotemporal resolution of the experiments (described in Sec.~\ref{sec:experiment}).
Therefore, we performed numerical simulations to clarify the relationship between the jet pressure impulse and the penetration depth.

The remainder of this paper is structured as follows.
In Sec.~\ref{sec:experiment}, the experimental setup and conditions of laser-induced microjet ejection and penetration are presented.
We also present an overview of microjet injection, the results of estimating the velocity distribution of the jet, and the variation of the penetration depth with respect to the standoff distance from the initial meniscus.
Then, in Sec.~\ref{sec:simulation}, we explain the method and present the results of the numerical simulations.
In addition, the method for estimating the jet pressure impulse is presented, and the results of a comparison with the actual penetration depth are discussed.
Finally, the conclusions and outlook are presented in Sec.~\ref{sec:conclusion}.

\section{Experimental investigation}\label{sec:experiment}
In this section, we describe the experimental setup and the conditions used for laser-induced microjet ejection and penetration.
We provide a visual representation of the microjet injection, an estimation of the velocity distribution of the jet, and the results of tests of the variation of penetration depth with respect to the standoff distance from the initial meniscus.

\subsection{Experimental setup and conditions}\label{subsec:exp_setup}
\begin{figure}[t]
\centering
\includegraphics[width=1.0\columnwidth]{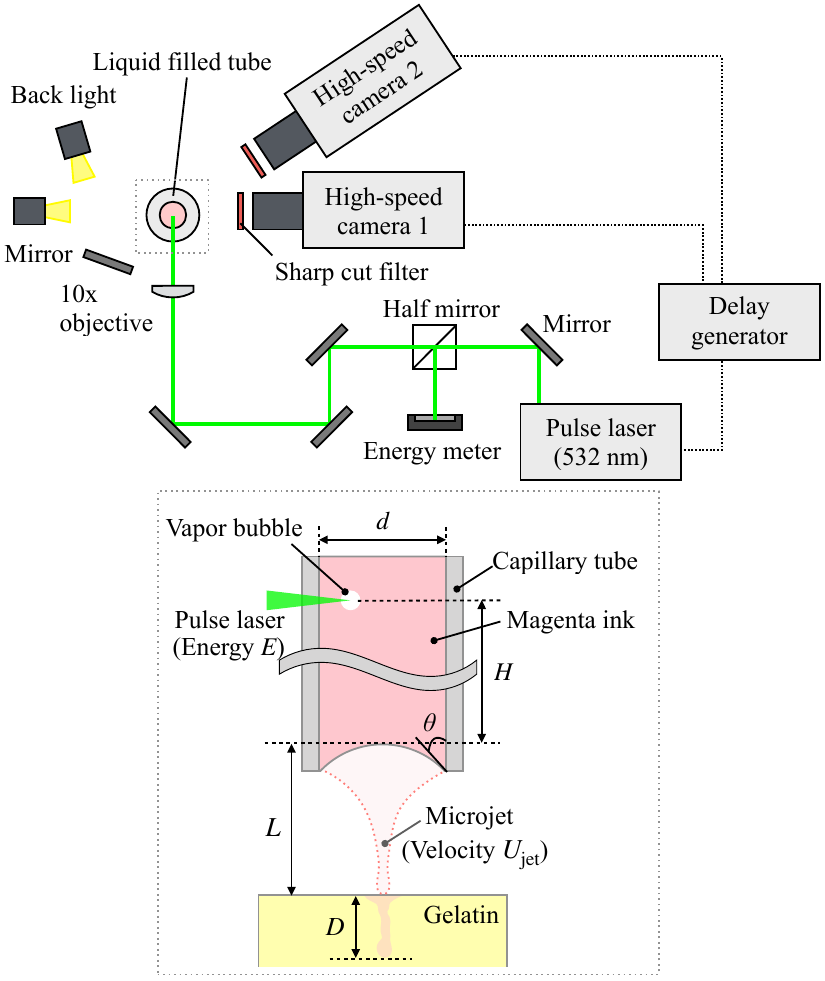}
\caption{\label{fig:exp_setup}Schematic drawing of the experimental setup for the ejection and penetration of a laser-induced microjet.}
\end{figure}

The experimental setup is shown in Fig.~\ref{fig:exp_setup}.
In this system, a capillary tube facing downward is connected to a syringe and filled with liquid.
Magenta ink (THC-7M4N, Elecom Co., Japan, viscosity: 2.7~mPa$\cdot \rm{s}$, density: 1\,060~$\rm{kg/m^3}$, surface tension: 40~mN/m), which has high energy-absorption efficiency for green light, is used as the liquid.
A transparent plastic container filled with a soft material is then held against the tube.
Gelatin with a concentration of 5\,wt.\,\%, which simulates human subcutaneous tissue\cite{menezes2009shock}, is used as the soft material.

A microjet is generated by focusing a pulsed laser (wavelength: 532~nm, pulse width: 6~ns) through an objective lens (MPLN10x, Olympus Co., Japan, magnification: 10$\times$, N.A. value: 0.25) onto the liquid in the tube.
By making the meniscus surface concave, a focused microjet is generated via the focusing effect\cite{tagawa2012highly, peters2013highly}.
A half-mirror (OptoSigma Co., transmission: 50\%) splits the pulsed laser into light directed toward the circular tube and light directed toward an energy meter (EnergyMax-RS J-10MB-HE, Coherent Co., USA) to measure the laser energy $E$.
Two high-speed cameras (FASTCAM SA-X, Photron Co., Japan, temporal resolution: 75\,000--150\,000~frames per second) are used to capture the view of the microjet ejection and penetration: Camera~1 captures the microjet penetration, and Camera~2 captures an overall image of the microjet being generated and injected into the soft material from an oblique angle that does not interfere with Camera~1.
The backlight of Camera~2 is projected via a mirror.
A sharp cut filter (SCF-50S-56O, Sigma Koki Co., Japan) is attached to the lens to block the 532-nm laser beam.
The two high-speed cameras and a laser generator (Nd:YAG laser; Nano S PIV, Litron Laser Co., USA) are synchronized by a delay generator (Model 575 Pulse/Delay Generator, BNC Co., USA).

In this study, we investigated the variation of the penetration depth $D$.
We varied the jet shape as a key factor by changing the standoff distance from the meniscus to the gelatin $L$ and the inner diameter of the microtube $d$, as shown in Table~\ref{tab:exp_param}.
The jet shape is considered to become thicker as the inner diameter of the microtube is increased.
It is known that the jet velocity $U_{\rm{jet}}$ has a significant effect on the penetration depth\cite{baxter2006needle, taberner2012needle}.
Therefore, to investigate this relationship, it is desirable to perform each experiment with an approximately constant jet velocity.
Tagawa \textit{et~al.}\cite{tagawa2012highly} showed that the jet velocity can be expressed as
\begin{equation}
\label{eq:Ujet}
U_{\rm{jet}} \simeq C_0\frac{(E-E_{\rm{heat}})(1+B \cos \theta)}{Hd},
\end{equation}
where $E$ is the laser energy measured by the energy meter, $E_{\rm{heat}}$ is the jet-generation threshold for the laser energy, $H$ is the distance between the laser's focal point and the initial meniscus, $\theta$ is the contact angle between the inner wall and the air--water interface, and $C_0$ and $B$ are fitting parameters.

Since Eq.~(\ref{eq:Ujet}) shows that $H$ and $d$ are inversely proportional to each other, we varied $H$ in response to $d$ to maintain a constant jet velocity.
However, since the energy-absorption efficiency varies with the inner diameter of the tube due to its curvature, the laser energy $E$ was adjusted to satisfy the target jet velocity of 70~m/s, which is a sufficient velocity for the jet to penetrate the gelatin. 
The measured jet velocity was found to have a certain range, as shown in Table~\ref{tab:exp_param}. 
This is because $\theta$ also had an error with a range of $\theta=0$--45° in each experiment.
Note that the error in the values is large due to the image resolution, and there are variations in the jet velocity even with the same laser energy.

In these experiments, the value of $U_{\rm{jet}}$ was estimated from the displacement of the jet tip in two consecutive images.
The images selected were the first image in which the jet tip appeared outside the capillary and the next image in the sequence.
Based on the time resolution of this experiment, we considered that the jets had reached near asymptotic and constant values in the first image.
However, under conditions where $L$ is small ($d=1.0$~mm: $L<0.9$~mm, $d=0.75$~mm: $L<0.9$~mm, $d=0.5$~mm: $L<0.9$~mm, $d=0.3$~mm: $L<0.7$~mm, $d=0.25$~mm: $L<0.4$~mm), the time taken for the jet tip to impact the gelatin is shorter than that between frames.
In these cases, $U_{\rm{jet}}$ was estimated from $E$, which is reported to have a linear relationship with the jet velocity\cite{tagawa2012highly}.
These relationships are approximations obtained from 40 microjet injection experiments, each with the same conditions.
We obtained eight $L$ conditions for each $d$ and performed six experiments for each condition.

\begin{table}[t]
\centering
\caption{\label{tab:exp_param}Experimental conditions for microjet injection: $d$ is the inner diameter of the tube, $L$ is the standoff distance between the initial meniscus and the gelatin, $H$ is the distance from the laser's focal point to the meniscus, $E$ is the target laser energy, and $U_{\rm{jet}}$ is the measured jet velocity.}
\begin{tabular}{ccccccccc}
\hline\hline
$d$ [mm] & $L$ [mm] & $H$ [mm] & $E$ [mJ] & $U_{\rm{jet}}$ [m/s]\\
\hline
1.00 & 0.50--5.00 & 4.0 & 2.0 & 68.5 $\pm$ ~3.1\\
0.75 & 0.50--5.00 & 3.0 & 1.6 & 75.9 $\pm$ ~2.9\\
0.50 & 0.30--5.00 & 2.0 & 0.5 & 73.3 $\pm$ ~2.7\\
0.30 & 0.30--4.00 & 1.0 & 0.2 & 69.4 $\pm$ ~6.9\\
0.25 & 0.25--3.00 & 1.0 & 0.1 & 81.0 $\pm$ ~5.7\\
\hline\hline
\end{tabular}
\end{table}

\subsection{Experimental results and discussion}\label{subsec:exp_RandD}
\subsubsection{Laser-induced microjet ejection and penetration}\label{subsec:microjet}
\begin{figure}[t]
\centering
\includegraphics[width=1.0\columnwidth]{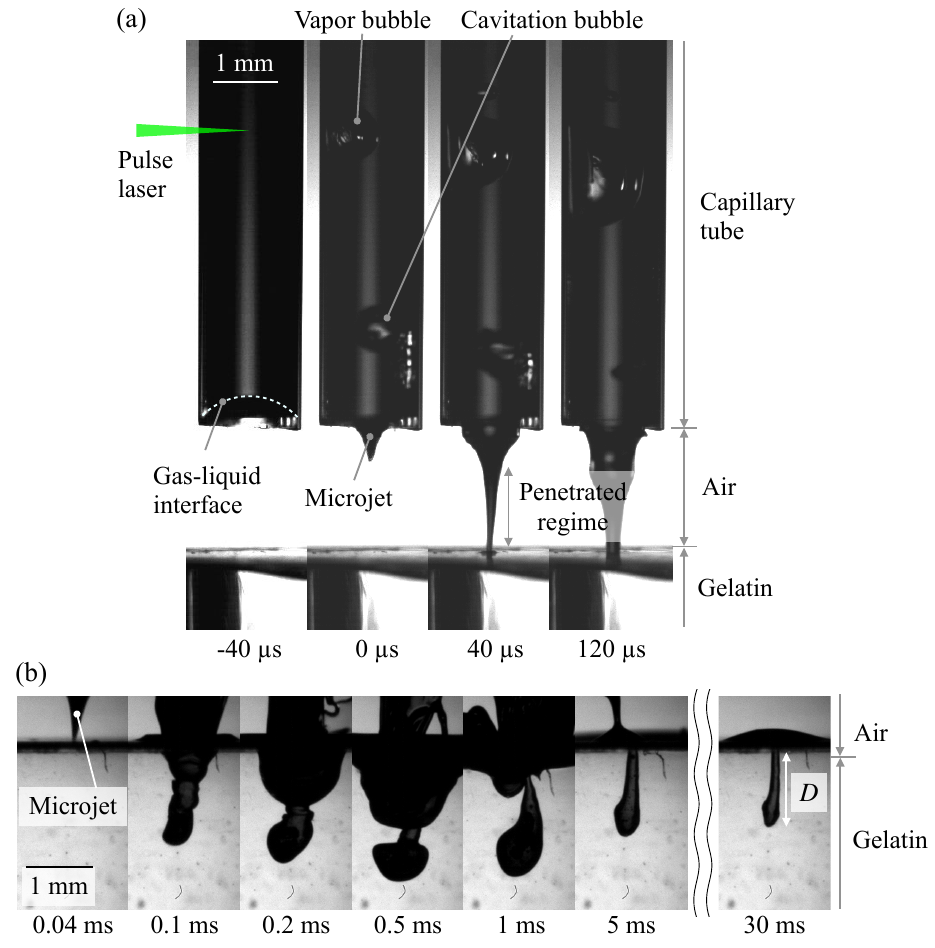}
\caption{\label{fig:microjet}Image sequence of (a)~ejection and (b)~penetration for the laser-induced microjet; the penetration depth $D$ is defined as the depth at $t=30$~ms.}
\end{figure}

Figure~\ref{fig:microjet} shows an example image sequence taken from the two cameras described above: the time evolution of the jet ejection is shown in Fig.~\ref{fig:microjet}(a), and the jet penetration is shown in Fig.~\ref{fig:microjet}(b).
In Fig.~\ref{fig:microjet}(a), we set the first frame in which the jet tip appears outside the capillary as $t=0$~$\mu$s. The meniscus surface is formed with a concave shape [Fig.~\ref{fig:microjet}(a): $t=-40$~$\mu$s], and the focused microjet is generated by absorption of the laser energy by the liquid [Fig.~\ref{fig:microjet}(a): $t=0$--$120$~$\mu$s].
At this time, a vapor bubble is generated at the laser's focal point due to a sudden temperature rise caused by laser absorption\cite{akhatov2001collapse, padilla2014optic}.
Then, secondary cavitation bubbles are generated by the sudden pressure drop in the liquid due to the propagation of expansion waves\cite{caupin2006cavitation, ida2009multibubble}.
After being ejected from the capillary, the focused microjet impacts the gelatin while maintaining its shape.

In Fig.~\ref{fig:microjet}(b), it can be seen that when the microjet impacts the gelatin, its tip forms a cavity.
The following region of the tip---which we define as the ``penetrated regime''---pushes through the cavity and enters the gelatin by causing it to deform elastically (0.1--0.5~ms).
The cavity is then pushed back by the elastic force of the gelatin after the maximum penetration (0.5~ms).
After a certain period of time, the depth of penetration finally stabilizes.
This phenomenon has also been observed in previous studies\cite{kiyama2019visualization, tagawa2013needle, schramm2004jet, stachowiak2009dynamic}.
In this experiment, $D$ is defined at $t=30$~ms, when the motion of the liquid has become sufficiently stable in all cases.

\begin{figure}[t]
\centering
\includegraphics[width=1.0\columnwidth]{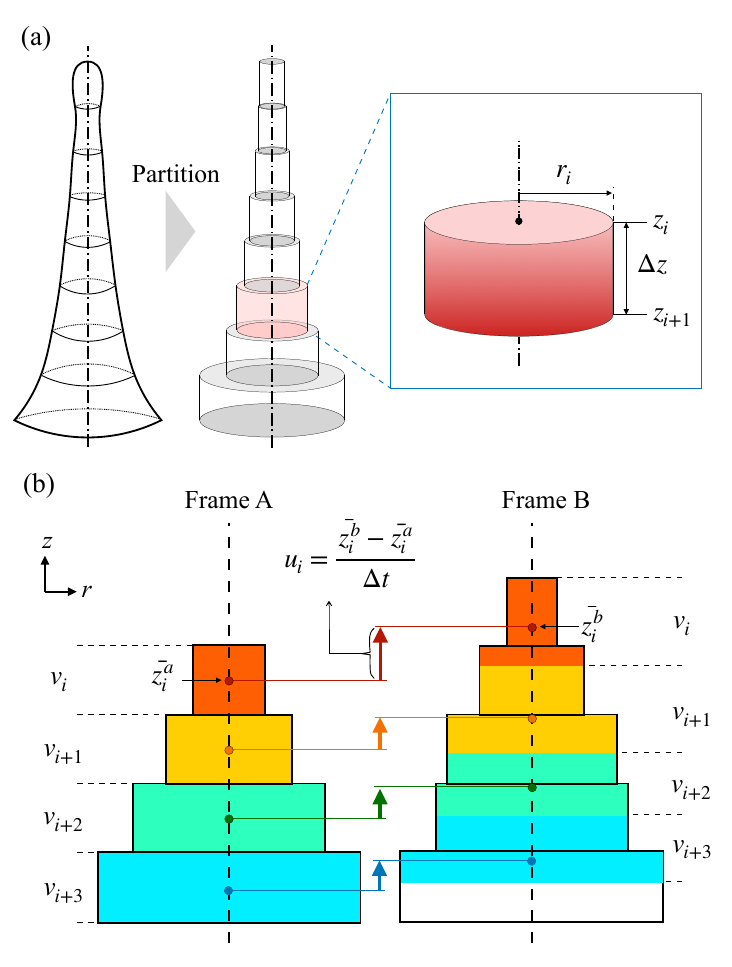}
\caption{\label{fig:exp_method}(a)~Schematic illustrations showing the modeling of the microjet as a stack of cylinders and dividing it into volume elements and (b)~estimating the velocity distribution $u_i$ using these volume elements.}
\end{figure}

\subsubsection{Velocity distribution of jet in the experiment}\label{subsec:exp_vel}
To estimate the jet pressure impulse, the velocity distribution in the $z$ direction of the microjet, which includes not only the tip velocity but also the velocity of the penetrated regime, is examined in this section.
In this work, the velocity distribution was estimated from the distances traveled by the centroids of each volume element, following the method of previous research\cite{van2014velocity}.
Note that the microjet is assumed to be an incompressible and axisymmetric flow.
As shown in Fig.~\ref{fig:exp_method}(a), the jet shape is assumed to be a stack of cylinders, divided into 3-pixel volume elements in the $z$ direction.
The volume elements $v_i$ are estimated from the radius $r_i$ and the height $\Delta z$ of each element:
\begin{equation}
\label{eq:vi}
v_i = \pi r_i^2 \Delta z.
\end{equation}
Here, to examine the $z$ direction, the velocity is obtained by tracking the centroid displacements of each element.
As shown in Fig.~\ref{fig:exp_method}(b), we compare Frame~A with Frame~B after $\Delta t$ seconds from Frame~A.
For the same volume estimated in Frame~A, we find the corresponding volumes in Frame~B, starting from the tip of the microjet.
From these volumes, we find the centroid of element $i$ in Frame~A, $\bar{z^a_i}$, and those of the corresponding volumes in Frame~B, $\bar{z^b_i}$.
Finally, for the corresponding elements in frames~A and B, the velocities $u_i$ are estimated from the centroid displacements:
\begin{equation}
\label{eq:ui}
u_i = \frac{\bar{z^b_i}-\bar{z^a_i}}{\Delta t}.
\end{equation}

\begin{figure}[t]
\centering
\includegraphics[width=1.0\columnwidth]{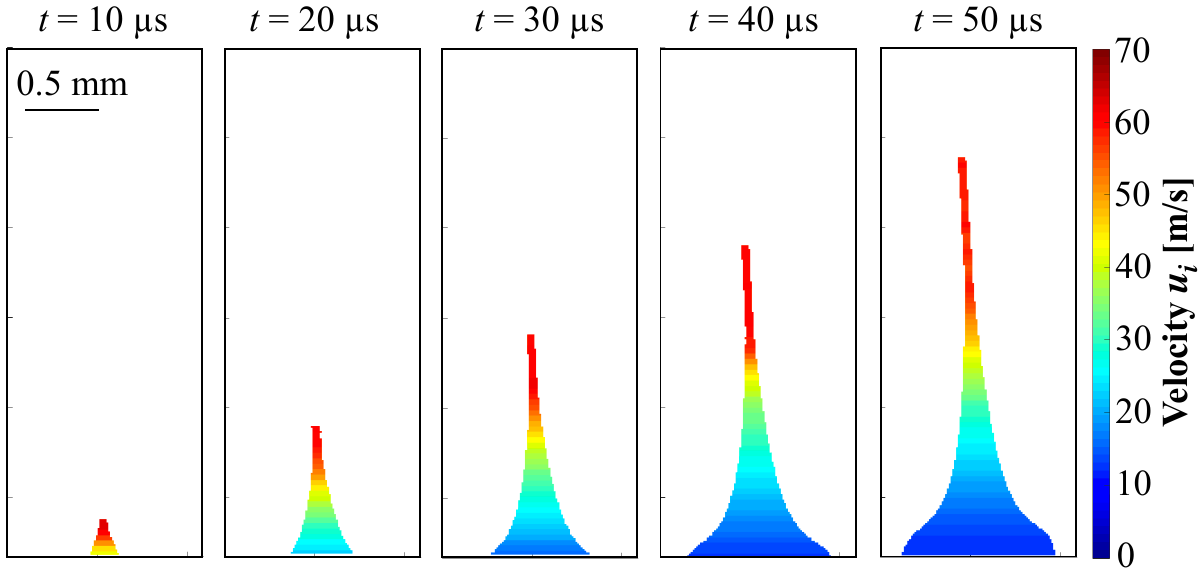}
\caption{\label{fig:exp_microjet}Color maps of the velocity distributions $u_i$ obtained from the experimental images for the case of inner diameter $d=1.0$~mm.}
\end{figure}

\begin{figure}[t]
\centering
\includegraphics[width=1.0\columnwidth]{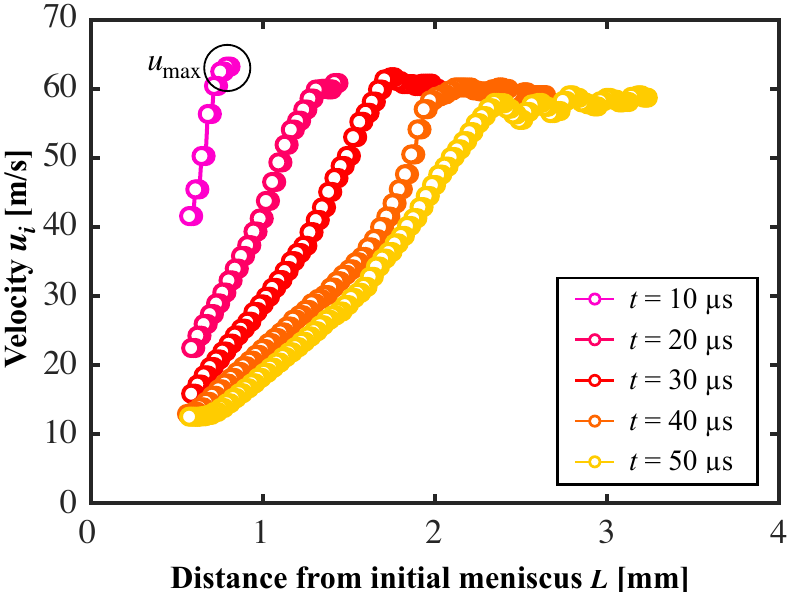}
\caption{\label{fig:v_L_exp}Velocity distributions with respect to standoff distance from the initial meniscus for the case of inner diameter $d=1.0$~mm.}
\end{figure}

Figure~\ref{fig:exp_microjet} shows a color map of the velocity distributions in the $z$ direction estimated using the above method.
The results are shown for $d=1.0$~mm, which had the thickest jet and the smallest error in estimating the velocity distribution in the experiments.
The time range is $t=10$--$50$~$\mu$s, which is within the measurable range.
The velocities are higher toward the jet tip and lower toward the jet root.
Figure~\ref{fig:v_L_exp} shows the velocity distribution as a function of the distance the microjet travels from the initial gas--liquid interface.
We can use these figures to describe the relationship between the jet velocity and the jet shape.

In the range of these measurements, the velocity at the jet root slowed sharply from about 40 to 10~m/s.
Correspondingly, the velocity gradient decreases as the jet progresses.
The velocity gradient near the jet tip is approximately zero, where the velocity remains nearly constant at about 60~m/s.
Considering the color map in Fig.~\ref{fig:exp_microjet}, it can be seen that this region has an elongated shape, and is considered to have velocity only in the $z$ direction.
However, during the thinning process of the jet shape, the fluid moves in the $r$ direction.
This indicates that the convergence of the fluid and the thinning of the jet result in $z$-velocity acceleration.
Since a high $z$ velocity is essential for the microjet to penetrate into the gelatin, this result suggests that the focusing shape is important for the penetration phenomenon.

\subsubsection{Penetration depth and standoff distance from initial meniscus}\label{subsec:D_L_expexp}
\begin{figure}[t]
\centering
\includegraphics[width=1.0\columnwidth]{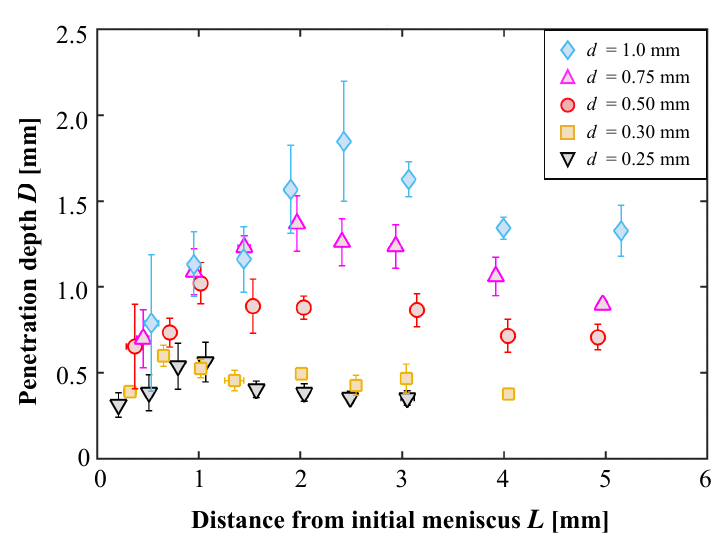}
\caption{\label{fig:D_L_exp}Relationship between the penetration depth $D$ and the standoff distance from the initial meniscus $L$ for all cases. The plotted values indicate the mean values derived from six experiments under the same conditions, with error bars representing one standard deviation of the uncertainty.}
\end{figure}

The relationship between the penetration depth $D$ and the standoff distance from the initial meniscus $L$ is shown in Fig.~\ref{fig:D_L_exp}.
The plots show the mean values of six experiments under the same conditions, and each of the error bars shows one standard deviation of the uncertainty.
Remarkably, this figure shows that $D$ tends to first increase and then decrease with respect to $L$ for all inner-diameter values.
Moreover, the peak position of the distance, defined as $L_{\rm{peak}}$, tends to be larger for larger $d$ values.
We argue that this is because the larger the inner diameter, the longer it takes to reach the focused shape, and the farther away the peak position $L_{\rm{peak}}$.

In the following, we describe the relationship between $d$ and $L_{\rm{peak}}$.
According to Tagawa \textit{et~al.}\cite{peters2013highly}, the timescale of flow focusing $\Delta t_{\rm{f}}$ can be estimated by
\begin{equation}
\label{eq:timescale}
\Delta t_{\rm{f}} \sim \frac{R}{U_0},
\end{equation}
where the radius of curvature at the gas--liquid interface is $R$ as a representative length and the initial velocity of the meniscus surface is $U_0$.
By assuming that the peak value of the penetration depth is obtained when the flow focusing is completed, $L_{\rm{peak}}$ is estimated as
\begin{equation}
\label{eq:Lpeak}
L_{\rm{peak}} \sim U_{\rm{f}} \Delta t_{\rm{f}}.
\end{equation}
Here, $U_{\rm{f}}$, which corresponds to the maximum velocity, is expressed as
\begin{equation}
\label{eq:UfU0}
U_{\rm{f}} = (1+\alpha \cos \theta)U_0,
\end{equation}
where $\alpha$ is a fitting parameter, and the radius of the curvature $R$ is expressed as
\begin{equation}
\label{eq:Rd}
R = \frac{d}{2\cos \theta}
\end{equation}
by assuming that the gas--liquid interface is spherical.
From Eqs.~(\ref{eq:timescale})--(\ref{eq:Rd}), the peak position $L_{\rm{peak}}$ is estimated as
\begin{equation}
\label{eq:Lpeakd}
L_{\rm{peak}} \sim \frac{1+\alpha \cos \theta}{2\cos \theta}d = \beta d,
\end{equation}
where $\beta$ is a fitting parameter.
From this equation, it can be seen that $L_{\rm{peak}}$ is proportional to $d$.
According to previous research\cite{peters2013highly}, $\alpha \sim2.0$ in this microjet, which means that $1.50<\beta<1.71$ in the range $0^{\circ}<\theta<45^{\circ}$.

\begin{figure}[t]
\centering
\includegraphics[width=0.9\columnwidth]{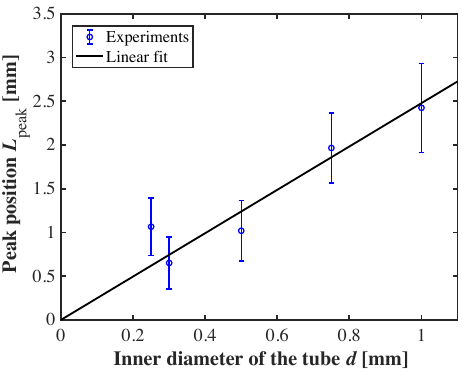}
\caption{\label{fig:d_Lpeak} Relationship between the peak position $L_{\rm{peak}}$ and the inner diameter of the tube $d$. Each of the error bars shows one standard deviation using the plots on the peak positions and either side of it in Fig.~\ref{fig:D_L_exp}.}
\end{figure}

The relationship between $L_{\rm{peak}}$ and $d$ obtained from Fig.~\ref{fig:D_L_exp} is shown in Fig.~\ref{fig:d_Lpeak}.
This plot shows the mean values of the peak positions from six experiments.
Each of the error bars shows one standard deviation using the plots on the peak positions and either side of it in Fig.~\ref{fig:D_L_exp}.
This plot demonstrates the proportional relationship between $d$ and $L_{\rm{peak}}$ according to Eq.~(\ref{eq:Lpeakd}).

The approximate fit line drawn in Fig.~\ref{fig:d_Lpeak} has a slope of 2.48, which is larger than the range of $\beta$ as calculated from Eq.~(\ref{eq:Lpeakd}).
This result seems to indicate that the assumption made in this section---that $L_{\rm{peak}}$ is obtained when the flow focusing is complete---is incorrect, because the actual peak values are significantly larger than our assumption.
In fact, defining the maximum value of the velocity elements as $u_{\rm{max}}$, the position of $u_{\rm{max}}$ with respect to $L$ is approximately $L=0.8$~mm ($t=10$~$\mu$s) for $d=1.0$~mm, as shown in Fig.~\ref{fig:v_L_exp}, while the peak penetration depth is approximately $L=2.4$~mm, as shown in Fig.~\ref{fig:D_L_exp}, which is inconsistent.

In light of the above, to find the critical factor affecting the penetration depth, we expect that it is necessary to consider the jet shape including the penetrated regime, which becomes more deformed as $L$ increases.
For this, we need to consider the jet pressure impulse, which is defined as the impulsive force applied per unit area, as will be further described in Sec.~\ref{subsec:sim_analize}.
However, the jet pressure impulse is difficult to estimate due to the limited spatiotemporal resolution of the experiments, as shown in Fig.~\ref{fig:v_L_exp}.
Therefore, we need numerical calculations to solve this problem.

\section{Discussion aided by numerical simulations: Jet pressure impulse}\label{sec:simulation}
This section describes the method and results of the numerical calculations performed to overcome the limitations of the experiments.
The numerical results are used to estimate the jet pressure impulse of a focused microjet.
The jet pressure impulse is compared with the penetration depth, and their relationship is discussed.
Note that the numerical simulations focus only on the jet-ejection process.

\subsection{Numerical calculations}\label{subsec:sim_method}
\begin{figure}[t]
\centering
\includegraphics[width=1.0\columnwidth]{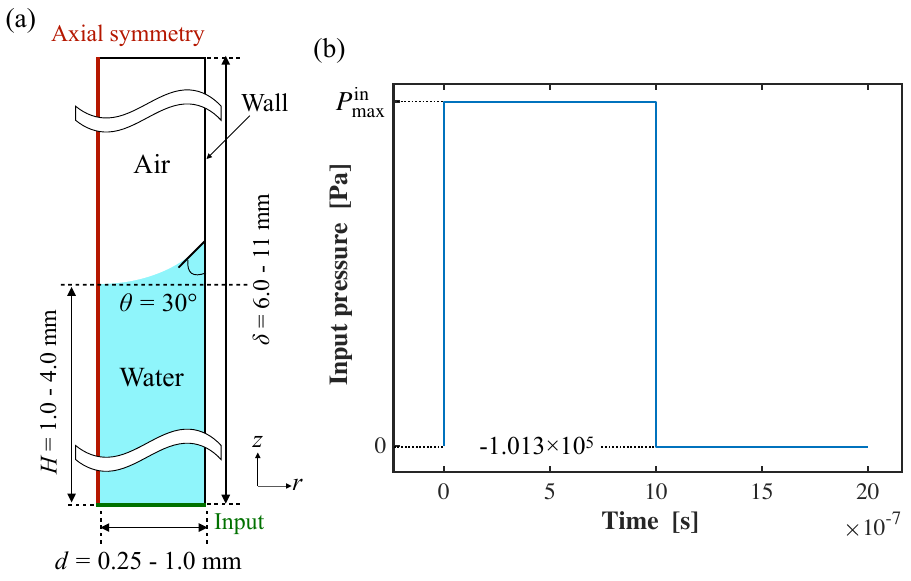}
\caption{\label{fig:sim_method}(a)~Geometry for the numerical simulations of microjet ejection with a two-dimensional axisymmetric model. (b)~Time evolution of the input pressure: starting with a peak pressure at $t=0$~$\mu$s, this pressure is maintained for $10$~$\mu$s, and it is subsequently pushed back down to atmospheric pressure.}
\end{figure}

This section describes the computational conditions for simulating the ejection process of a liquid (pure water) from a capillary as a focused jet using commercial software (COMSOL Multiphysics).
The governing equations are the incompressible Navier--Stokes equation and the continuity equation, which are defined as
\begin{equation}
\label{eq:Navier}
\rho \left[ \frac{\partial \mathbf{u}}{\partial t} + (\mathbf{u} \cdot \nabla) \mathbf{u} \right] = - \nabla p_{\rm{N}} + \mu \nabla^2 \mathbf{u},
\end{equation}
\begin{equation}
\label{eq:continuity}
\nabla \cdot \mathbf{u} = 0,
\end{equation}
where $\rho$ is the density of water (998~$\rm{kg/m^3}$) and $\mu$ is its viscosity (1.0~mPa$\cdot \rm{s}$), and $p_{\rm{N}}$ is the pressure of the liquid.
The geometry to be calculated is shown in Fig.~\ref{fig:sim_method}(a).
This is a two-dimensional axisymmetric model with the center of the tube (left side of the domain) as the axis.
The liquid is filled into the tube with a concave air--water interface.
As in the experiments, the inner diameter of the tube $d$ was varied, being given values of 0.25, 0.30, 0.50, 0.75, and 1.00~mm.
The height of the computational domain $\delta$ was varied as shown in Table~\ref{tab:sim_param} to set the domain to sufficiently calculate over the range of the distance $L$ in the experiments.
The distance from the bottom of the tube to the air--water interface $H$ was correspondingly varied by the same values as in the experiments.
To obtain the same flow-focusing effect, the contact angle was assumed to be constant at $\theta=30^\circ$---for which a clear focusing jet shape is observed---in all cases.

\begin{table}[t]
\centering
\caption{\label{tab:sim_param}Numerical conditions for the microjet ejection. Here, $d$ is the inner diameter of the tube, $\delta$ is the height of the computational domain, $H$ is the distance from the laser's focal point to the meniscus, $U_{0}$ is the initial velocity of the meniscus surface.
}
\begin{tabular}{ccccccccc}
\hline\hline
$d$ [mm] & $\delta$ [mm] & $H$ [mm] & $U_0$ [m/s] \\
\hline
1.00  & 6.0 & 4.0 & 20.14 \\
0.75 & 7.0 & 3.0 & 20.00 \\
0.50 & 9.0 & 2.0 & 22.40 \\
0.30 & 10.0  & 1.0 & 24.80 \\
0.25 & 11.0  & 1.0 & 25.95 \\
\hline\hline
\end{tabular}
\end{table}

Using the above computational domain, the boundary conditions of pressure at the bottom of the domain were input as a driving force similar to that provided by the laser in the experiments.
The time evolution of the input pressure is shown in Fig.~\ref{fig:sim_method}(b).
The input pressure is given a maximum value at $t=0$~$\mu$s; it is then held at that pressure for $10$~$\mu$s before being reduced again to atmospheric pressure (assuming that the vapor pressure is significantly smaller than the atmospheric pressure).
Note that the maximum pressure $P^{\rm{in}}_{\rm{max}}$ can be estimated by
\begin{equation}
\label{eq:Pinput}
P^{\rm{in}}_{\rm{max}} = \frac{\rho U_0 H}{\Delta t^{\rm{in}}},
\end{equation}
where $\Delta t^{\rm{in}}$ is the time span over which the input pressure is applied, which is $10$~$\mu$s.
As shown in Table~\ref{tab:sim_param}, $U_0$ was changed for each diameter to obtain the similar $U_{\rm{jet}}$ value as in the experiments.
The pressure at the top of the domain was set to zero.
A no-slip wall boundary condition was imposed on the right-hand side of the domain.
A square mesh with a maximum size of 1.4\%--1.5\% of the inner diameter was used as the computational grid.

\begin{figure}[t]
\centering
\includegraphics[width=1.0\columnwidth]{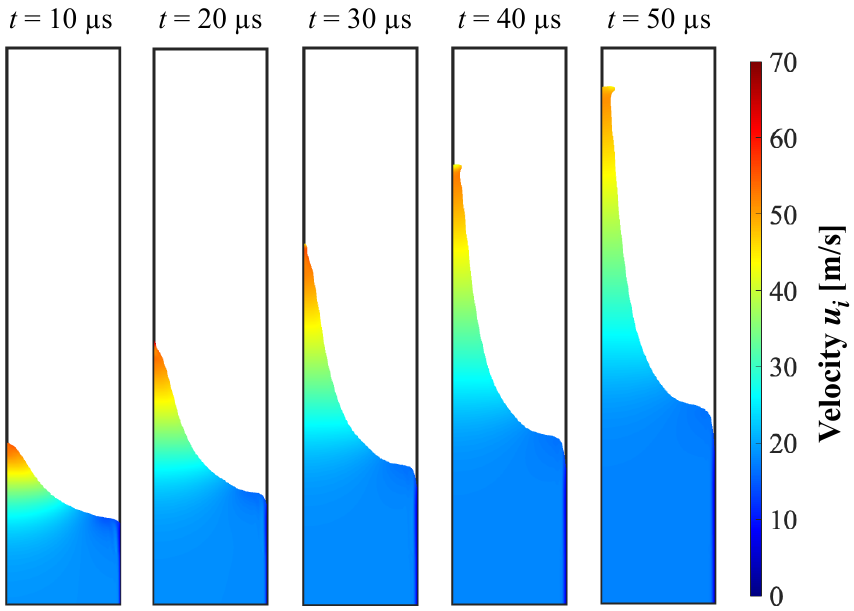}
\caption{\label{fig:sim_microjet}Color maps of the velocity distributions $u_i$ derived from the numerical calculations for the case of inner diameter $d=1.0$~mm.}
\end{figure}

\begin{figure*}[!t]
\centering
\includegraphics[width=1.7\columnwidth]{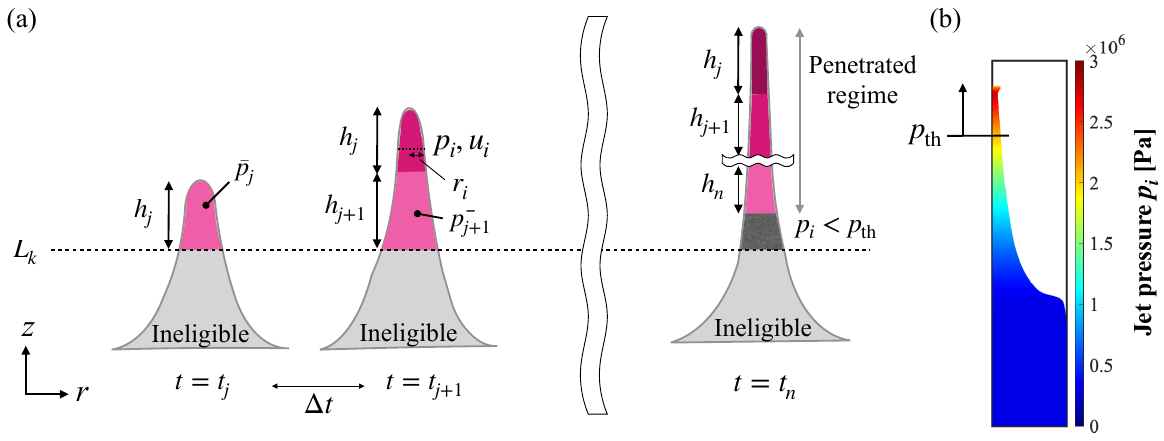}
\caption{\label{fig:Pjet_method}(a)~Schematic diagram of the jet pressure impulse estimation method. Jet pressure impulse is estimated from the jet pressure that exceeds a certain threshold $p_{\rm{th}}$ and has passed through length $L_k$. (b)~Distribution of jet pressure $p_i$ for the case $t=50$~$\mu$s with $d=1.0$~mm.}
\end{figure*}

The level set method \cite{sussman1994level,sussman1998improved, tomar2005numerical, li1993study}, an interface-detection method, was used to track the gas--liquid interface.
In this method, the gas--liquid interface is obtained by solving the advection equation for the level set function $\phi$:
\begin{equation}
\label{eq:levelset}
\frac{\partial \phi}{\partial t} + \mathbf{u} \cdot \nabla \phi = \gamma \nabla \cdot \left(\epsilon \nabla \phi - \phi(1-\phi) \frac{\nabla \phi}{|\nabla \phi|}\right),
\end{equation}
where $\gamma$ is the amount of reinitialization and $\epsilon$ is the interface thickness.
Herein, $\phi=1$ indicates the liquid phase, $\phi=0$ indicates the gas phase, $\phi=0.5$ indicates the gas--liquid interface, and $0<\phi<1$ indicates the mixed layer.
In the mixed layer, $\rho$ and $\mu$ are calculated as
\begin{equation}
\label{eq:levelset_rho}
\rho = \rho_{\mathrm{l}} \phi + \rho_{\mathrm{g}} (1-\phi),
\end{equation}
\begin{equation}
\label{eq:levelset_mu}
\mu = \mu_{\mathrm{l}} \phi + \mu_{\mathrm{g}} (1-\phi),
\end{equation}
where the subscripts $\mathrm{l}$ and $\mathrm{g}$ denote the liquid and gas phases, respectively.

The velocity distribution for the $d=1.0$~mm case of the numerical simulation results is shown in Fig.~\ref{fig:sim_microjet}.
When compared with the experimental velocity distribution shown in Fig.~\ref{fig:exp_microjet}, it can be seen that the velocity gradient from the tip to the bottom of the jet is similar, with a velocity of about 60~m/s at the tip and about 20~m/s at the bottom of the jet.
However, although the jet shape is thinner toward the tip, as in the experiments, the jet tip in the numerical simulations is rounded, which is different from the experimental observations.
The liquid jet is stretched, and droplets are then torn from the tip (pinched off)\cite{gordillo2020impulsive, shinjo2010simulation, leppinen2003capillary, burton2005scaling}.
The rounded area at the tip indicates the region where this pinch-off eventually occurs. 
Due to the limitation of the spatial resolution, the jet tip in the experiment is not rounded.
From this analysis, we expect that the numerical calculations reproduce the actual phenomenon, and it is thus valid to discuss the jet pressure impulse using these numerical results.

\subsection{Estimation method: Jet pressure impulse}\label{subsec:sim_analize}

\begin{figure*}[!t]
\centering
\includegraphics[width=2.0\columnwidth]{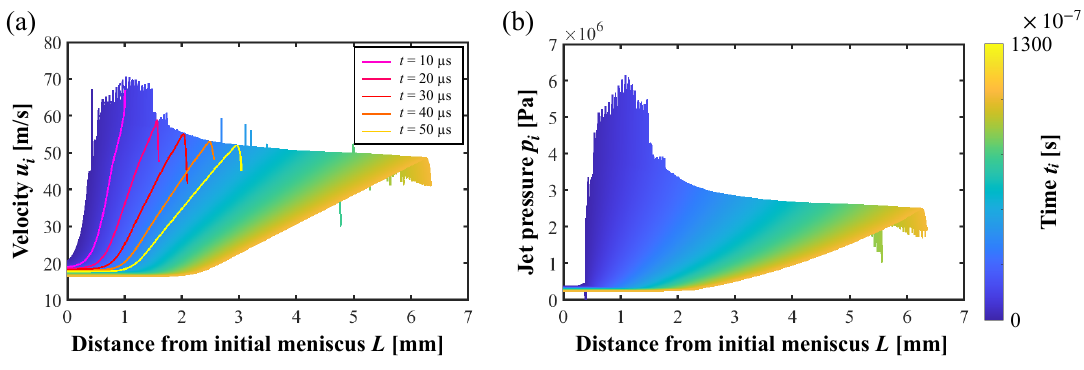}
\caption{\label{fig:v_Pjet_L} Distributions of (a)~velocity $u_i$ and (b)~jet pressure $p_i$ with respect to the standoff distance from the initial meniscus $L$. The different colors indicate time, and the velocity distribution at specific times as in the experiment ($t=10$--$50$~$\mu$s) are also highlighted in panel (a). Both results show the case of $d=1.0$~mm.}
\end{figure*}

We estimated the jet pressure impulse using the velocity distribution obtained numerically by the method described in the previous section.
A schematic diagram of the jet pressure impulse estimation method is shown in Fig.~\ref{fig:Pjet_method}(a).
First, we estimate the jet pressure at each cross section $p_{i}(z)$.
This is described as
\begin{equation}
\label{eq:pi}
p_{i}(z)=\frac{2\rho \int_0^{r_{i}(z)} u_{i}(z,r)^2r\mathrm{d}r}{r_{i}(z)^2},
\end{equation}
where $r_i$ and $u_i$ are the radius and $z$ velocity of the jet at each cross section $i$.
Note that Fig.~\ref{fig:Pjet_method}(b) shows the distribution of $p_i$ for the case $t=50$~$\mu$s with $d=1.0$~mm.
A discussion of this distribution is presented in the next section.
Since $r_i$ is variable in the $z$ direction, we need to estimate the average jet pressure at each time $\bar{p_j}$.
This is expressed as
\begin{equation}
\label{eq:pibar}
\bar{p_j} = \frac{\int^{h_j} p_{i}(z)r_{i}(z)^2\mathrm{d}z}{\int^{h_j}{r_{i}(z)^2}\mathrm{d}z}.
\end{equation}
Note that
\begin{equation}
\label{eq:hj}
h_j=u^{\rm{tip}}_{i} \Delta t,
\end{equation}
where $u^{\rm{tip}}_{i}$ is the tip velocity of each element.
As shown in Figs.~\ref{fig:Pjet_method}(a) and \ref{fig:Pjet_method}(b), we only estimate the jet pressure that exceeds a certain threshold $p_{\rm{th}}$ and has passed through the length $L_k$, which is defined as a discrete value every 0.1~mm in each range of $L$ in Table~\ref{tab:exp_param}.
From this, the pressure that can exceed the yield point at a distance $L$ from the initial meniscus is used for the estimation.

The threshold is set at $p_{\rm{th}}=2.35 \times 10^6$~Pa.
This was found to be the optimal value after tuning.
Assuming that the velocity distribution of the jet is constant in the $r$ direction, $p_{\rm{th}}=\rho u_{\rm{th}}^2$ is obtained from Eq.~(\ref{eq:pi}), where $u_{\rm{th}}$ is the velocity threshold.
Estimating $u_{\rm{th}}$ from this equation, it is found that $u_{\rm{th}}=48.5$~m/s for $p_{\rm{th}}=2.35 \times 10^6$~Pa.
In an experiment conducted by Tagawa \textit{et~al.}\cite{tagawa2013needle}, to penetrate 5\,wt.\,\% gelatin with a similarly focused microjet, it was observed that the jet failed to penetrate at $u_i<53$~m/s.
Based on this, it can be said that the selected threshold value is not unusual.
However, further understanding of the physical meaning of this value is needed for practical applications.

We finally estimated the jet pressure impulse as summed in time $\Pi_k$:
\begin{equation}
\label{eq:Pik}
\Pi_k = \sum^{n-1} \bar{p_j}\Delta t,
\end{equation}
where $\Delta t$ was set to $1.0 \times 10^{-7}$.
Moreover, $n$ was set as the frame at which $p_i$ first satisfies $p_i<p_{\rm{th}}$.
Thus, the jet pressure impulse is defined as the sum of the pressures from the time at which the jet tip first exceeds $L_k$ to that when there is no pressure above the threshold $P_{\rm{th}}$.
The above method allows us to estimate the pressure impulse from the tip to the penetrated regime of the jet at any distance $L$.

\subsection{Numerical calculation results}\label{subsec:sim_result}
Figures~\ref{fig:v_Pjet_L}(a) and \ref{fig:v_Pjet_L}(b) respectively show the distribution of the velocity $u_i$ and the jet pressure $p_i$ obtained from Eq.~(\ref{eq:Pinput}) with respect to $L$ for the case of $d=1.0$~mm.
Note that the color bar indicates the time elapsed since the jet was generated.
The specific velocity distributions at the same times as in the experiments are also highlighted in Fig.~\ref{fig:v_Pjet_L}(a).

The trend of the velocity distribution is generally similar to that of the experiment during $t=10$--$50$~$\mu$s (Fig.~\ref{fig:v_L_exp}).
Moreover, the jet pressure distribution is generally proportional to the square of the jet velocity.
Therefore, the trend between the jet velocity and the jet pressure does not change significantly.
The numerical calculations also show that the focused shape of the jet tends to accelerate the $z$ velocity, resulting in large values immediately after the jet is generated, as in the experiments.
However, the numerical calculations increase the spatiotemporal resolution in comparison to the experimental results, so the peak positions are clearly visible in the figure.
Note that there is some noise in the figure due to computational errors.
In addition, the position of the maximum velocity and the tip velocity are different from those found in the experiments.
As for the position of the maximum velocity, we argue that the error in the velocity distribution increases toward the tip due to the lack of spatial resolution in the experimental estimation method for the velocity distribution.
Regarding the tip velocity, in addition to the reasons just outlined, the jet tip is rounded in the numerical calculations, resulting in a sharp decrease in the jet velocity.
Therefore, we need to consider not only the jet tip but also the penetrated regime.
This is estimated by considering the jet pressure impulse.

\begin{figure*}[!t]
\centering
\includegraphics[width=0.92\textwidth]{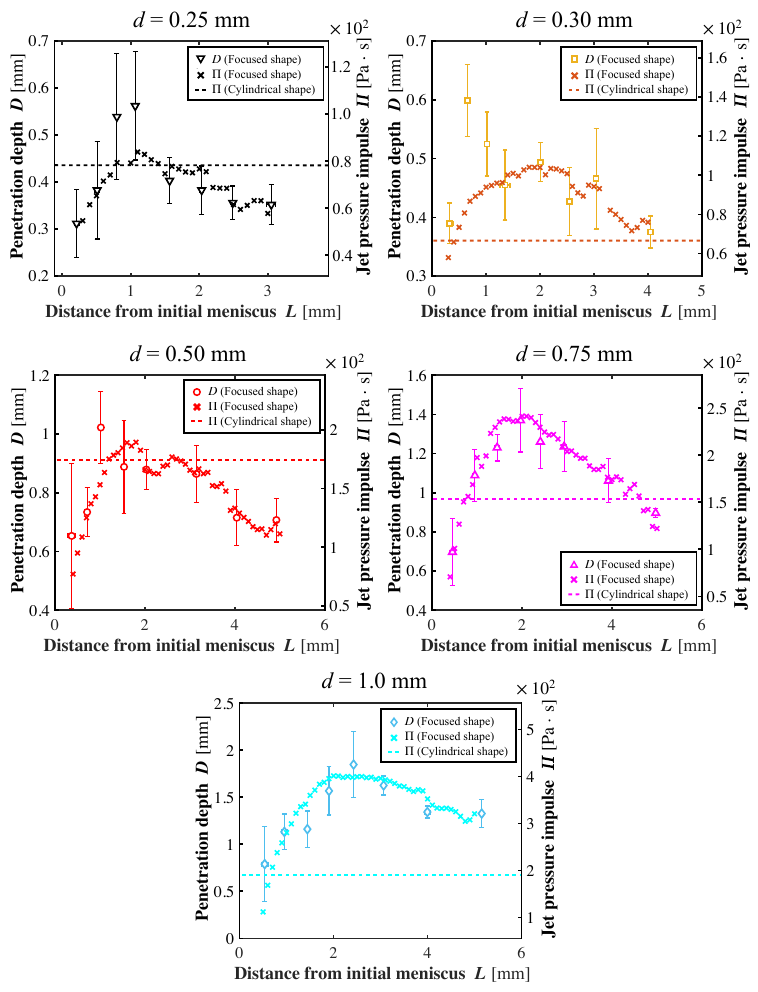}
\caption{\label{fig:D_L_sim}Penetration depth $D$ (left axis) and jet pressure impulse $\Pi$ (right axis) for the focused jet and the cylindrical jet with respect to the standoff distance from the initial meniscus $L$ in each inner diameter.}
\end{figure*}

The jet pressure impulse is estimated using the method described in the previous section.
Figure~\ref{fig:D_L_sim} shows the penetration depth $D$ (left axis) and the jet pressure impulse $\Pi$ (right axis) with respect to $L$ at each $d$.
As shown in the panels, the magnitude of the jet pressure impulse has a strong relationship with the jet penetration depth in all cases.
This indicates that the jet penetration depth can be estimated from the jet pressure impulse.
Moreover, peak values are observed in all cases due to the focused shape of the jet.
To illustrate this, let's consider a hypothetical situation where the jet is cylindrical. 
In this cylindrical jet scenario, $r_i(z)=u_i(z,r)=const.$ in Eq.~(\ref{eq:pi}), (\ref{eq:pibar}) gives $p_i=\bar{p_i}=\rho u_i^2$.
Therefore, from the equation:
\begin{equation}
\label{eq:n_frame}
n = \frac{1}{\Delta t} \frac{h_{\rm{c}}}{u_i},
\end{equation}
where $h_{\rm{c}}$ is the height of the penetrated regime for the cylindrical jet, we derive the following equation:
\begin{equation}
\label{eq:Pik_circ}
\Pi_k = \sum^{n-1} \rho u_i^2\Delta t \simeq \rho h_{\rm{c}} u_i.
\end{equation}
This equation shows that the jet pressure impulse of a cylindrical jet is independent of $L$, which is inconsistent with the experimental results, where the penetration depth exhibits peak values.
It is clearly confirmed in Fig.~\ref{fig:D_L_sim} that shows the pressure impulse of the cylindrical jet.
Note that the diameter of the cylindrical jet is 75\% of $d$, and the volume is the average of the total volume used in Eq.~(\ref{eq:Pik}) for each $L$ calculated for the focused jet.
For example, for a cylindrical jet at $d=1.0$ mm, the diameter $d_{\rm{c}}$ is 0.75 mm, the height $h_{\rm{c}}$ is 2.73 mm and the volume $v_{\rm{c}}$ is 1.21 mm$^3$.
This volume is consistent with the order $O(10^0)$ mm$^3$ at $t=0.5$ ms, which corresponds to the maximum penetration observed in the experimental result (Fig.~\ref{fig:microjet}).
Consequently, for a focused jet, the pressure impulse exhibits peak values, aligning with the penetration depth measurements.
These findings contribute significantly to the control of the penetration depth of focused microjets.

\clearpage

\section{Conclusions and outlook}\label{sec:conclusion}
The objective of this study was to elucidate a key parameter determining the optimal standoff distance with respect to penetration depth for the practical application of a needle-free injector using a high-speed focused microjet.
For this purpose, experiments examining microjet injection into a soft material were performed by varying the inner diameter of the tube.

First, the velocity distribution of the jet was estimated from high-speed images of the microjet injection.
The results show that the direction-of-travel component of the velocity distribution is larger near the tip because the jet is more focused in this region.
Furthermore, the maximum velocities have a peak value with respect to the standoff distance from the initial meniscus $L$.
The experiments also showed that the penetration depth has such a peak value with respect to $L$ for all tube diameters.
However, these peak positions are inconsistent.
We expect that this is because not only the velocity but also the shape and the penetrated regime of the jet are important for the penetration depth.
Therefore, we considered finding a way to estimate the jet pressure impulse, a physical parameter that can take into account the shape of the jet including the penetrated regime.
However, it was difficult to estimate the jet pressure impulse from the experimental results due to a lack of spatiotemporal resolution.

Thus, we used numerical calculations to simulate the same situations as in the experiments to estimate the jet pressure impulse.
The numerical simulations successfully generated jets with velocity distributions similar to those observed in the experiments.
The jet pressure impulse---i.e., the integrated values of the jet pressure passing a certain distance $L$ and exceeding a certain jet pressure---was estimated from the simulation results.
Remarkably, the jet pressure impulse and the penetration depth showed peak positions that coincided with respect to $L$.
Furthermore, the peak positions moved away from the meniscus as the inner diameter increased.
This is because the larger the inner diameter of the tube, the longer it takes for the jet to focus.
It is suggested that the optimal distance can be estimated by considering the jet pressure impulse.
This indicates that there is a correlation between the jet pressure impulse and the penetration depth.
These results show that the jet pressure impulse is a key parameter determining the penetration depth of the focused microjet.
The findings from this study are important for controlling the penetration depth of focused microjets.

In the future, we will consider the applicability of these findings by changing the hardness of the penetrated object to consider additional practical uses.
In the present estimation method, the hardness of the gelatin can be expressed by the jet-pressure threshold.
Therefore, we will explore the versatility of the present estimation method by experimentally and numerically examining the relationship between these two parameters.
Since the hardness of human skin varies from person to person \cite{agache1980mechanical, pailler2008vivo, kiyama2019gelatine}, this knowledge will provide further insights for practical applications.
Based on the above plan, we aim to achieve advanced control of the penetration depth of a high-speed focused microjet and to attain practical needle-free injection into human skin using this microjet system.

\begin{acknowledgments}
This work was funded by the Japan Society for the Promotion of Science (Grant Nos. 20H00223, 20H00222, and 20K20972), the Japan Science and Technology Agency PRESTO (Grant No. JPMJPR21O5), and Japan Agency for Medical Research and Development (Grant No. JP22he0422016).
The authors would also like to thank Dr. Masaharu Kameda (Professor, Tokyo University of Agriculture and Technology) for helpful discussions and comments.
\end{acknowledgments}

\section*{Author declarations}
\subsection*{Conflict of Interest}
The authors have no conflicts to disclose.

\section*{Data availability}
The data that support the findings of this study are available from the corresponding author upon reasonable request.

\section*{References}

\nocite{*}
\bibliography{aipsamp}

\end{document}